\documentclass[sigconf, screen]{acmart}
\AtBeginDocument{%
  \providecommand\BibTeX{{%
    \normalfont B\kern-0.5em{\scshape i\kern-0.25em b}\kern-0.8em\TeX}}}

\copyrightyear{2024}
\acmYear{2024}
\setcopyright{rightsretained}
\acmConference[CSCW Companion '24]{Companion of the 2024 Computer-Supported Cooperative Work and Social Computing}{November 9--13, 2024}{San Jose, Costa Rica}
\acmBooktitle{Companion of the 2024 Computer-Supported Cooperative Work and Social Computing (CSCW Companion '24), November 9--13, 2024, San Jose, Costa Rica}\acmDOI{10.1145/3678884.3681885}
\acmISBN{979-8-4007-1114-5/24/11}

\usepackage{graphicx}
\usepackage{booktabs}

\begin{document}


\title{Envisioning Possibilities and Challenges of AI for Personalized Cancer Care}

\author{Elaine Kong}
\affiliation{%
  \institution{University of Pittsburgh}
  \streetaddress{Pittsburgh}
  \city{Pittsburgh}
  \country{USA}}
\email{elainekong@pitt.edu}

\author{Kuo-Ting (Tim) Huang}
\affiliation{%
  \institution{University of Pittsburgh}
  \streetaddress{Pittsburgh}
  \city{Pittsburgh}
  \country{USA}}
\email{timhuang@pitt.edu}

\author{Aakash Gautam}
\affiliation{%
\institution{University of Pittsburgh}
  \streetaddress{Pittsburgh}
  \city{Pittsburgh}
  \country{USA}}
\email{aakash@pitt.edu}







\begin{abstract}





The use of Artificial Intelligence (AI) in healthcare, including in caring for cancer survivors, has gained significant interest. However, gaps remain in our understanding of how such AI systems can provide care, especially for ethnic and racial minority groups who continue to face care disparities.
Through interviews with six cancer survivors, we identify critical gaps in current healthcare systems such as a lack of personalized care and insufficient cultural and linguistic accommodation. AI, when applied to care, was seen as a way to address these issues by enabling real-time, culturally aligned, and linguistically appropriate interactions. 
We also uncovered concerns about the implications of AI-driven personalization, such as data privacy, loss of human touch in caregiving, and the risk of echo chambers that limit exposure to diverse information. 
We conclude by discussing the trade-offs between AI-enhanced personalization and the need for structural changes in healthcare that go beyond technological solutions, leading us to argue that we should begin by asking, ``Why personalization?''

\end{abstract}




\maketitle

\section{Introduction}

Cancer survivors, especially those from ethnic and racial minority groups, face profound challenges in their recovery journey, often exacerbated by significant disparities in accessing adequate resources for information and care \cite{nikkhah2022family, suh2020parallel, song2022design}. 
For instance, research shows that as many as half of all cancer survivors suffer from mental health issues such as anxiety, depression, and the fear of their cancer returning \cite{smith2024feasibility}. 
Despite the prevalence of these issues, psychosocial needs are often overlooked within oncological care, signaling a critical area for improvement \cite{mitchell2013depression, tauber2019effect}.

In this mix, with the advent of Artificial Intelligence (AI), there is a growing interest in exploring AI as a solution to bridge these gaps. 
As a tool, AI may have some potential.
To have any potential role, AI systems need to attend to and align with, the complex realities surrounding cancer survivors.
Scholars widely argued for a comprehensive approach to health by considering physical, emotional, social, and functional factors \cite{khan2023no, delpierre2023precision}.
Understanding the humans for whom the purported AI solution is designed, is thus essential for AI to have any possibility of success.

This exploratory work aims to take a step in this direction.
We conducted semi-structured interviews with six cancer survivors to uncover the psychosocial needs specific to this group and evaluate the potential of AI-based interventions to address these needs.
Our findings reveal a critical demand for social support, personalized care, and culturally sensitive resources that current healthcare infrastructures struggle to fulfill, which led the participants to envision possibilities of AI in the space.
Building on these findings, we demarcate the potential role and challenges of AI technologies in supporting the well-being of cancer survivors.

\section{Information-Seeking Behavior During Cancer Journey}

Each cancer journey is profoundly individual, and care services need to attend to these personal realities \cite{smriti2023bringing, purohit2023chatgpt, jacobs2014cancer}. 
Scholarship highlights several barriers in accessing services \cite{ghiotti2023prototyping, nikkhah2022family, suh2020parallel}, as well as the critical challenge in supporting access to relevant information \cite{kamita2020promotion}. 
In this space, research identifies two predominant information-seeking behaviors among cancer patients: active seeking and avoidance \cite{case2005avoiding, eheman2009information}.
The stigmatization often associated with cancer can influence patient behaviors, with some individuals withdrawing and others actively seeking information \cite{fourie2015feminist}. These behaviors are critical because they shape patient-provider interactions and significantly impact treatment outcomes. Scholars argue that healthcare strategies must adapt to individual emotional states and information preferences to improve these interactions and ensure more effective treatment \cite{sankaran2019enhancing, bertrand2023selective}.


The issue is of relevance to the HCI and CSCW community since digital platforms are increasingly leveraged by cancer survivors and caregivers to gain health information \cite{chen2013caring, burgess2022care, jacobs2014cancer}. 
It is estimated that approximately 45\% of cancer patients use digital platforms for information \cite{hesse2005trust}. 
Scholars have argued for personalized technology solutions that cater to the unique circumstances and needs of cancer survivors and their caregivers \cite{nikkhah2022feel, suh2020parallel}, supporting the aims of identifying gaps in current technologies and proposing innovative design ideas \cite{randazzo2023trauma, del2023sound}.
Moreover, the exploration of health information-seeking behavior provides critical insights into the complexities of patient care and self-management, particularly for those navigating chronic diseases like cancer \cite{song2022design, zghab2024s}.

We live in an era of AI Realism where AI is pervasive \cite{gautam2024reconfiguring}.
There is a growing trend of developing AI solutions for healthcare, including supporting cancer survivors \cite{sebastian2022artificial, parimbelli2021review}. 
In particular, AI-based solutions are positioned as promising tools, including in improving access to personalized information \cite{chanda2021mindnotes, murnane2018personal, tarver2019use}. 
We examine some of the needs and build on the findings to chart out the scope of possibilities and dangers in leveraging AI to support cancer survivors.

\section{Data Collection and Analysis}


\begin{table*}[]
\caption{Participant demographics. All the data are self-reported.}
\vspace{-1em}
\label{tab:my-table}
\resizebox{\textwidth}{!}{%
\begin{tabular}{llllllllc}
\hline
\multicolumn{1}{c}{\textbf{ID}} & \multicolumn{1}{c}{\textbf{\begin{tabular}[c]{@{}c@{}}Age \\ Range\end{tabular}}} & \multicolumn{1}{c}{\textbf{Gender}} & \multicolumn{1}{c}{\textbf{Race/ Ethnicity}} & \multicolumn{1}{c}{\textbf{Cancer Type}} & \multicolumn{1}{c}{\textbf{\begin{tabular}[c]{@{}c@{}}Stage at \\ Diagnosis\end{tabular}}} & \multicolumn{1}{c}{\textbf{Current Stage}} & \multicolumn{1}{c}{\textbf{\begin{tabular}[c]{@{}c@{}}Treatment(s) \\ so far\end{tabular}}} & \multicolumn{1}{c}{\textbf{\begin{tabular}[c]{@{}c@{}}Confidence with AI\\ (1-5; 5=`very confident')\end{tabular}}} \\ \hline
P1 & 45-54 & Female & \begin{tabular}[c]{@{}l@{}}Black or African \\ American\end{tabular} & Breast & Stage II & In remission & \begin{tabular}[c]{@{}l@{}}Chemotherapy, \\ Surgery\end{tabular} & 3 \\ \hline
P2 & 45-54 & \begin{tabular}[c]{@{}l@{}}Prefer not \\ to say\end{tabular} & \begin{tabular}[c]{@{}l@{}}I describe myself \\ in some other way\end{tabular} & Melanoma & Stage II & Stage I & Radiation therapy & 4 \\ \hline
P3 & 25-34 & Male & Asian & Prostate & Stage II & In remission & Surgery & 4 \\ \hline
P4 & 55-64 & Female & Asian & Lung & Stage II & In remission & \begin{tabular}[c]{@{}l@{}}Chemotherapy, Radiation \\ therapy, Surgery\end{tabular} & 3 \\ \hline
P5 & 45-54 & Male & Asian & Lung & Stage I & In remission & Surgery & 3 \\ \hline
P6 & 65+ & Female & Asian & Colorectal & Stage I & In remission & Chemotherapy, Surgery & 3 \\ \hline
\end{tabular}%
}
\end{table*}


As part of our pilot study, we interviewed six cancer survivors from diverse backgrounds, ensuring a representation of varied cancer types and stages (see Table \ref{tab:my-table}). 
The semi-structured interviews were between 45 to 90 minutes long. 
Interview questions explored participants’ information needs and their experiences with healthcare services.
We also discussed the participants’ sources of community support and their perceptions of using AI in their cancer care journey.
All interviews were conducted remotely considering the participants' preferences and geographical constraints. 
We recorded the audio.

We conducted thematic analysis of the transcript, starting from open-coding and iteratively working towards higher-level codes. 
Following Salda{\~n}a's guidelines for ``solo coding'' \cite{saldana2021coding}, the first author, who did the open coding alone, consulted and discussed the emerging codes with the last author throughout the coding process.  
Intermediate categories that emerged included ``Experience with Current Interventions,'' ``Unique Needs and Challenges,'' and ``Perception and Openness to AI-based Solutions.'' 
Through discussions and iterative grouping, we arrived at four themes that form the basis of our findings below.
We conducted member checking to ensure an authentic representation of participants' ideas.

\section{Findings}

A salient aspect in our conversations was around the lack of information and care tailored to the participants' unique situation in their journey.
The unmet informational needs and the lack of personalized healthcare services impacted their well-being. To navigate this, participants relied on online resources, which provided a community for support but also brought misinformation. 
All of these situated their vision of AI's possibilities and challenges in providing care.

\subsection{Unmet Informational Needs for their Unique Situation}

While the informational needs among the six participants varied, there was a wide consensus on the need for transparency about their condition and the possible pathways for treatment. 
For instance, participant P6 voiced her concern because of the uncertainty, stating that she was initially directed to a specific type of chemotherapy without adequate information on other available treatments, \textit{"that I likely did not receive enough information about the type of chemo, or other available chemo options."}
In fact, five participants reported an increased anxiety about their health and mortality, noting heightened concerns about their overall health following the diagnosis. 

In the face of a lack of transparency from care providers, people often rely on information online.
Four participants shared how they specifically searched for information about the possibility of their cancer returning. As P1 shared, \textit{``Considering that many studies indicate a 60\% chance of recurrence, could it come back?''} 
There is significant misinformation and conflicting perspectives online \cite{swire2020public}. 
As P2 lamented, \textit{``I think I don’t want to Google anymore, sometimes the information is not accurate and not reliable. Finding reliable information online is hard; there's too much conflicting advice.''}
Four of the six participants noted that misinformation had negatively impacted their emotional well-being during their recovery journey, \textit{``The quality of information and the rapid pace at which it is disseminated have had significant health-related impacts. The spread of unreliable evidence about cancer care has amplified treatment hesitancy, and I am very worried about the increased tension and anxiety it causes.''}. 
Some expressed increased anxiety and fears about recurrence and general health after encountering misleading information, \textit{``I fear that my cancer will come back. I am depressed and feel as worried as I did when I was first diagnosed with cancer.''} 

Participants such as P1 and P5 also engaged in online groups to gain and share information. 
The participant, P5 highlighted the value of an online cancer support network for his specific type of cancer, mentioning how networks would share support and sometimes meet in person, \textit{``The networks provide an invaluable connection. I've learned some great stress relief techniques and I've made some wonderful friends through them. Whether our meetings are virtual or face-to-face, they remind us that we're not alone.''}. 
While overall positive, some participants reported that the online groups did not meet their informational needs, as heard in P4's account, \textit{``I joined a cancer survivor support group online. The support groups were helpful, but sometimes the topics didn’t quite match my needs.''}

\subsection{Need for Personalized Healthcare Services}

The participants shared varied experiences with healthcare professionals, noting positive interactions when providers dedicated time to offer personalized care.
P2, for example, recalled a particularly supportive nurse, \textit{``she took the time to really listen. It made all the difference.''}
Recollecting positive experiences, participants often focused on instances when healthcare providers inquired about the participants' emotional needs, which the participants felt was affirming, P4 noted \textit{``Providers often demonstrated empathy by expressing their understanding of my difficult circumstances. It sounds like it’s really affecting my state of mind.''} 
These points suggest a desire for personalized care tailored to individual needs.

Not all interactions with healthcare providers were positive, often centering around the lack of personalized care. 
Participants shared feeling neglected or misunderstood by healthcare professionals.  
Five of the six participants explicitly mentioned a significant lack of psycho-social support from healthcare providers.
For instance, P1 highlighted the lack of emotional support that met them where they were, \textit{``I felt lost in the sea of generic advice. I needed something that spoke to my experience, to what I was going through.''}  
A participant, P4, shared feeling anxious and depressed during their cancer journey, \textit{ ``I feel stressed, frightened, and panicky. I don't have an appetite, and I don't want to go out or see others socially.''}.
The lack of personalized care that acknowledges their unique journey with cancer was acutely felt as can be heard in P3's expression, \textit{``I was just another case to him, nothing more.''}
Despite the growing popularity of patient-centered approaches, factors such as the burden on healthcare providers and the impersonal nature of technology-mediated interactions hinder personalized care.
P1 noted that the hospital system felt overwhelmed, which made personalized care challenging, \textit{``One of the big problems is that our hospital system is getting crushed. Now it's that hospitals and medical practices all over the country are short-staffed. This makes it more difficult to implement personalized care.''}.

\subsection{Seeking Support From Community}

Social support from family and friends was a pivotal element for five participants, providing significant emotional relief during their treatment.
The participants shared about support from their family members, which provided comfort and reduced the isolation often felt during cancer treatment. For example, P3 appreciated the opportunity to share what he was going through with his loved ones, \textit{``It was good for me to discuss my side effects of the chemo with others ... so I could talk to my son about that, giving me a way to voice my experiences.''}

Some found support beyond family and friends. 
For example, P4 shared the benefits she got by reaching out to social workers through her work, \textit{``I work with many social workers. I was very comfortable talking with them about it ... It was therapeutic to talk about it, and it helped alleviate some of my anxiety as well.''}
Likewise, two participants emphasized the significance of receiving support from their religious communities. 
P1, for example, emphasized the important role her church members played in supporting her, \textit{``My church consistently shows its support, makes it really clear that they are there for us, praying for us, and they are watching over us.''}

Some participants also reached out online for support.
Two participants, P5 and P6 also reported reaching out to cancer survivors online for support both during and after their treatment. 
For example, P6 noted, \textit{``I have met a lot of people that have similar cancer experience, and they are willing to offer support and communicate.''} 
Similarly, P5 shared the value of going through the journey together, \textit{``I received a lot of support and was told that, having gone through a similar trauma, if others can survive, so can I. It may seem like there's no end in sight, but if I keep fighting, I can win too.''} 

In these instances, the participants were seeking information and support that aligned with their cultural practices. 
Our participants were from minoritized identities and felt they were outsiders, as we could hear in P1's account, \textit{``I didn’t feel that the services considered my cultural background, which affected how I received the support.''}
Five participants' experiences suggested the importance of culturally sensitive interventions that respect and address the distinct needs of various communities. 
For example, P5 noted, \textit{``It would be great if the healthcare provider had the same or a similar ethnic or religious background, as it can help lay the foundation for trust in the relationship. I appreciate being able to receive culturally sensitive care that respects my spiritual needs.''} 
Asian communities often rely on traditional methods for health, including cancer treatment. Participants like P3 felt that their ongoing services did not acknowledge traditional perspectives, \textit{``It’s hard to find culturally sensitive support that understands my specific community’s viewpoint on cancer.'' }

Similarly, participants expressed challenges in overcoming language barriers when interacting with healthcare providers, as heard in P6's account, \textit{``I'm facing a language barrier, so it would be perfect if the healthcare provider could help translate.''} 
P4 stated that she reached out to her family for help, \textit{``My English is not good, I need to ask my daughter to translate for me.''}
These accounts of participants reaching out to family and friends, communities near them, and online groups suggest the importance of social support. 
It also highlights the critical ways in which people are trying to navigate and access personalized support that meets them where they are in their cancer journey.

\subsection{Perceived Role of AI in Cancer Support}
Amidst the salient focus on personalized care and informational needs, it was natural that our discussion involved AI.
All participants reported that they had heard about AI but their exposure to AI was limited; P3 was the only participant who had used an AI-enabled app --- a chatbot --- for services.
Nonetheless, the participants valued the possibility that AI could provide support throughout the day, which would not be reasonable to expect of human caregivers.
P3, for example, shared \textit{``It } [AI] \textit{ could provide continuous support, which is hard to get from human services. I like the idea of an AI that checks in on me.''}
In our findings, we note both the possibilities and pitfalls of using AI in supporting them in their recovery and care. 

As highlighted above, participants faced several social and technical barriers to accessing personalized information and care. 
This led them to envision ways in which AI could potentially fill the gaps. 
When asked about how they thought AI could help, all participants talked about personalization to various degrees. 
P1 shared wanting to access support in tracking their health status, \textit{``I would like a symptom tracker and a way to learn more about managing side effects.'' }
Similarly, P3 wanted information about resources that were easy for him to understand, \textit{``Health resources that are easy to understand and access would help a lot.'' }
P2, likewise, wanted responsive immediate support through AI. 
Others like P5 and P6 were more general but clearly stated that their adaptation of AI would depend on how well it supports personalization, as heard in P5's account, \textit{``The more tailored the advice fit my needs, the more likely I am to use it.''}
P6, in fact, felt that personalization would enable trust, stating, \textit{``I believe personalization would make the difference in whether I trust and continue to use the AI.'' }

Beyond personalization, participants envisioned AI helping with broader healthcare and informational issues.
For example, P2 wanted \textit{``... to have a tool that regularly reminds me of health tips would help me manage my condition better.''}
We also learned about their vision of AI for mental health support, which highlighted the healthcare gaps they believed AI could address.
For example, P5 saw the potential of AI to help with anxiety, \textit{``AI tool could be helpful for my emotional through immediate responses and support to cope with anxieties.''}
P6 believed AI could help simulate interactions to handle depression, \textit{``For those of us dealing with depression, an AI tool that monitors mood changes and provides proactive interactions could be very useful.''}
Similarly, P1 shared the value of easily accessing AI as a simulation to talk for comfort, \textit{``The accessibility} [of AI] \textit{are definitely appealing. It can simulate talking to a real person anytime I need support is comforting.''}

While these accounts suggest optimism for AI in cancer care, participants also shared hesitation and concerns about the risks associated with AI systems. 
For example, P1 shared, \textit{``I’m open to it if it can provide anonymity and ease of access.''}. 
P4 too shared concerns about their privacy, \textit{``I worry about the privacy of my data and how it's used.''}
P3, on the other hand, wanted AI to acknowledge their traditional treatment approach, \textit{``Comfortable } [with it] \textit{as long as it complements traditional therapies and not replace them.''} 
Others were more skeptical of AI's capabilities and reliability as heard in P5's account, \textit{``My concern would be how well AI can truly understand human emotions.''}
Similarly, P6 shared, \textit{``I'm skeptical about whether AI can offer the same level of empathy as human interaction,''} highlighting the value placed on human connections and interactions for care services. 
\section{\vspace{-0.2em}Discussion and Conclusion}

In our interviews, cancer survivors shared their vision of an ideal healthcare service: supportive, inclusive, and responsive care that attends to their unique needs. 
Their experiences illuminated the lack of transparency and responsive support, which led participants to feel less agentic in their interactions with healthcare systems and in making decisions that affect their own treatments. 
These gaps, which arose from structural problems, led participants to envision better possibilities with AI.
The participants had limited knowledge of AI, but the ubiquity of AI had captured their attention. 
We heard accounts of the cancer survivors' vision for AI, particularly in aligning information and care with their cultural values and linguistic abilities, and accessing responsive care.

\subsection{Personalization in Cancer Care and AI}

A primary use case for AI involved aligning the care practices with the survivors' cultural practices. 
Cancer survivors have diverse cultural backgrounds and personalized AI was seen as a way to provide care that is attuned to cultural sensitivities \cite{conley2021multiple}.  For instance, P1 and P3 highlighted the importance of receiving culturally relevant support, noting that services that consider their cultural background would enhance their satisfaction. AI tools can be designed to recognize and adapt to diverse cultural backgrounds and health beliefs of patients \cite{nadarzynski2024achieving}. This not only enhances the patient experience but also improves adherence to treatment by respecting and integrating cultural practices and values into the care process \cite{hilty2021mobile}.

Related to this, participants shared their challenges in navigating foreign languages with their healthcare providers, raising challenges in receiving accurate information and expressing their health concerns. They envisioned inclusive care that accommodates the linguistic diversity of cancer survivors \cite{whitehead2023barriers}.
AI was seen as a tool to offer multilingual support and bridge communication gaps.

Similarly, participants envisioned possibilities for AI to enable responsive care anytime they need it. With human caregivers, this ability would not be affordable for most.  Visions of tailored treatment details and personalized management tips have been argued to improve the management of their condition \cite{ahmed2020artificial}.

\subsection{Case Against Personalization in Cancer Care}

Envisioning of AI in personalized healthcare also brought to light significant concerns. 
Participants worried about the implications of such personalization. 
Participants like P4 expressed apprehension about data privacy and how their information would be used. 
The growing attention to explainable and transparent AI can address these concerns (e.g., \cite{panch2019artificial, esmaeilzadeh2020use}). 
Moreover, the participants' accounts highlighted the importance of involving them in the decision-making process, creating avenues for them to be in dialogue with healthcare providers, designers, and other actors providing care. 

We also heard concerns about the possibility of losing human connections in care. 
Participants shared how they reached out to people for information and care.
They valued the human touch. 
With technology-centered approaches, they were concerned about potentially eroding this human touch in care, as we heard in P6's concerns with AI's lack of empathy.  

Participants also expressed concerns about technology imposing approaches that disregarded their cultural values and practices.
As we heard in the interviews, a significant thrust for demanding personalization was that they felt their cultural norms and values were not currently centered in their care.
In this respect, they sought information and care in their community, both online and offline.

\subsection{Tensions in Designing for Personalized Care}
In the participants' accounts, we heard a desire for personalized information and care. This was evident in their expressed needs for care and information aligned with their cultural values, the ability to interact with healthcare providers in their native languages, and access to responsive care whenever needed.

A cursory reading of these desires for personalized information and care suggests significant potential for AI in this space. AI systems can help healthcare providers learn about cultural practices and support real-time translation \cite{chen2020artificial}, enabling effective communication across different linguistic abilities \cite{liu2024computer}. Moreover, AI is often positioned as a tool to provide personalized assistance at any time.

We argue, however, that the desire for AI-driven care solutions reflects a deeper issue --- the need for control and agency in the face of a malady that erodes both \cite{mukherjee2010emperor}. The problem is structural. The increasing rate of cancer among the global population \cite{soerjomataram2021planning}, the rising cost of healthcare \cite{prager2018global}, and the general devaluation of care practices \cite{nikkhah2022family} are underlying problems that manifest as a lack of personalized care.

From our findings, we draw three major implications for personalization and AI in cancer care. 
The first implication revolves around the social elements to support patients in making informed decisions in a space that is rife with a lack of transparency and limited agentic involvement of the patients.
The second emerges from an appreciation of the community that is formed when people seek diverse sources of information, which could be in conflict when personalized information is centered. 
The third implication is more structural, where we believe there is a tension between the increasing push for AI in healthcare and the structural changes required to provide the holistic support patients need.

\subsubsection{Personalization and Informed Decision Making}

Promoting survivors' control and agency over technology used in cancer care is a critical consideration.
The participants shared concerns about a lack of transparency which highlighted the importance of individualized health literacy. 
Increasing health literacy, including about their condition and the possibilities and limitations of technology in care, is fundamental in enabling patients to make informed decisions about their treatments. 
This was echoed by participants who felt that more detailed and comprehensible information could have significantly improved their experiences. 

AI systems are complex because they make decisions using generalized models that lack context, though they can be adapted to include specific contextual information. 
This requires ensuring that AI systems are tailored to each case and that the patients know what it can and cannot do.
A form of critical literacy needs to be incorporated along with the introduction of AI for healthcare \cite{kimiafar2023artificial, jordan2010conceptualising}.  
Thus, while AI systems can be a tool to enable personalized care, those systems need to be made personalized first by empowering patients on its use and even the possibility to refuse its use.

\subsubsection{Personalization and the Risk of Echo Chambers}

In our findings, salient elements involved the participants' desire to gain access to diverse information about care possibilities. 
They sought information from multiple sources. 
This was a way for the participants to have control over the decisions they had to make. Having a community that shared similar experiences and included members whom they could reach out to for information and care was critical to this end.

The use of AI as a tool for personalized care may conflict with the need for diverse information sources. 
With personalization, patients could encounter information that is narrow; for example, only information that aligns with their existing views and preferences. 
This concern was highlighted by P2 and P6, who noted the importance of diverse information sources. 
Moreover, a sole focus on personalization through AI can erode the sense of community since people will have limited common ground around shared experiences.  
AI systems should be designed to introduce patients to a variety of perspectives and options, preventing bias reinforcement and promoting well-rounded, informed decision-making.
Efficient information delivery alone is not sufficient in this case.

\subsubsection{Personalization and Structural Change Trade-off}

As we reflect on our conversations, we note that participants initially focused on structural problems they faced in accessing information and care.
But their envisioned solution focused significantly on AI tools and their possibilities. 
We find this mismatch between the underlying issue and the envisioned solutions particularly problematic because it can potentially distract change-makers from engaging with the structural issues.  

The desire for personalization, at its core, involves an expression of wanting deep care that encompasses different aspects of patients' lives. 
Such care is becoming increasingly more expensive to afford. 
It is a privilege that is not easily accessible to the majority \cite{patel2020cancer,sullivan2011delivering}. 
Thus, the underlying problem is a structural one. 
AI as a technological tool can, at best, attend to some of those structural problems. There is a need for institutional and infrastructural approaches.
However, the primacy of AI solutions risks overlooking the urgent need to address the broader structural issues that impact cancer care and healthcare more generally.

\subsection{Implication: Ask ``Why Personalization?''}
Our findings highlight that personalization was viewed as a response to underlying issues encountered during cancer care, such as limited transparency and control over care practices, high costs of care services, and a lack of culturally sensitive support. We argue for the importance of asking ``Why personalization?'' to surface and engage with underlying issues \emph{before} designing for personalization.   

Asking ``why personalization?'' encourages undertaking a participatory approach, bringing the cancer care survivors into the design process.  
Participatory approach, \emph{when well designed}, can help cancer patients gain power and control over their care processes.
Critical health and technology literacy, making available broader sources of information and care, and situating technology in the context of structural issues surrounding healthcare are necessary conditions to support agentic participation in the design process. 

In the case of technology for personalization, the growing trend of involving people impacted by AI systems in the design process could provide a path forward. 
But echoing \citet{delgado2023participatory} and \citet{birhane2022power}, we urge scholars and practitioners aiming to engage in participatory AI to pay attention to who is involved and how they are involved. 
As our findings show, participants, even though they share some common elements of their identity (e.g., cancer patient), are not homogenous and have differing needs and desires. 
Similarly, systems of care are multi-faceted and complex. 
Each would require and desire different kinds and degrees of participation. 
Some may require \emph{not} using AI; an option that should be made available to people who are purported to be supported by the care infrastructure.

\begin{acks}
We are deeply thankful to the six participants for their time.
\end{acks}

\bibliographystyle{ACM-Reference-Format}
\bibliography{sample-base}


\end{document}